\documentclass[sigconf]{acmart}
\AtBeginDocument{%
  }

\begin{document}

\title{CGM-Led Multimodal Tracking with Chatbot Support: An Autoethnography in Sub-Health}

\acmConference[ICHEC 2025]{International Conference on Human-Engaged Computing (ICHEC 2025)}{November 21--23, 2025}{Singapore}
\acmYear{2025}
\acmBooktitle{Proceedings of the International Conference on Human-Engaged Computing (ICHEC 2025), November 21--23, 2025, Singapore}
\acmPrice{15.00}

\author{Dongyijie Primo Pan}
\email{dpan750@connect.hkust-gz.edu.cn}
\affiliation{%
  \institution{The Hong Kong University of Science and Technology (Guangzhou)}
  \city{Guangzhou}\state{Guangdong}\country{China}}

\author{Lan Luo}
\email{lluo476@connect.hkust-gz.edu.cn}
\affiliation{%
  \institution{The Hong Kong University of Science and Technology (Guangzhou)}
  \city{Guangzhou}\state{Guangdong}\country{China}}

\author{Yike Wang}
\email{yikewang@tongji.edu.cn}
\affiliation{%
  \institution{Tongji University}
  \city{Shanghai}\country{China}}

\author{Pan Hui}
\email{panhui@hkust-gz.edu.cn}
\affiliation{%
  \institution{The Hong Kong University of Science and Technology (Guangzhou)}
  \city{Guangzhou}\state{Guangdong}\country{China}}
\affiliation{%
  \institution{The Hong Kong University of Science and Technology}
  \country{Hong Kong SAR}}

\renewcommand{\shortauthors}{Pan et al.}

\begin{abstract}
Metabolic disorders present a pressing global health challenge, with China carrying the world’s largest burden. While continuous glucose monitoring (CGM) has transformed diabetes care, its potential for supporting sub-health populations—such as individuals who are overweight, prediabetic, or anxious—remains underexplored. At the same time, large language models (LLMs) are increasingly used in health coaching, yet CGM is rarely incorporated as a first-class signal. To address this gap, we conducted a six-week autoethnography, combining CGM with multimodal indicators captured via common digital devices and a chatbot that offered personalized reflections and explanations of glucose fluctuations. Our findings show how CGM-led, data-first multimodal tracking, coupled with conversational support, shaped everyday practices of diet, activity, stress, and wellbeing. This work contributes to HCI by extending CGM research beyond clinical diabetes and demonstrating how LLM-driven agents can support preventive health and reflection in at-risk populations.
\end{abstract}

\ccsdesc[500]{Human-centered computing~Empirical studies in HCI}
\ccsdesc[300]{Human-centered computing~Ubiquitous and mobile computing systems and tools}
\ccsdesc[300]{Applied computing~Health informatics}
\ccsdesc[100]{Human-centered computing~Natural language interfaces}

\keywords{continuous glucose monitoring (CGM), chatbot, personal informatics, autoethnography}

\maketitle

\section{Introduction}

Over the past decade, metabolic disorders have escalated worldwide, posing major challenges for public health systems~\cite{chew2023global}. In China, an estimated 148 million adults aged 20--79 currently live with diabetes—about \textbf{11\%} of the adult population, the highest national burden globally. This number is projected to reach 168.3 million by 2050, nearly \textbf{18\%} of adults~\cite{li2020prevalence,wang2017prevalence}. Parallel increases in overweight and obesity further heighten metabolic risks and psychological stress~\cite{Tamashiro2011Metabolic}. In this context, continuous glucose monitoring (CGM) has shifted from sporadic laboratory tests to continuous everyday tracking~\cite{rodbard2016continuous}, offering real-time curves, trend arrows, and metrics such as Time in Range (TIR)~\cite{danne2017international}. While CGM has reshaped diabetes care and data interpretation, existing clinical and HCI research has focused primarily on diagnosed populations such as type 1 and type 2 diabetes~\cite{10.1145/3613904.3642234,10.1145/3613904.3642435,10.1145/3706599.3719780,10.1145/3706598.3713247}. Yet medical evidence suggests that long-term glycemic variability is also present in overweight, prediabetic, and other sub-health groups, and such fluctuations are a critical precursor to diabetes onset~\cite{vigersky2019relationship}. This gap underscores the need for the HCI community to broaden its lens beyond clinical diabetes, exploring how CGM data and related multimodal tracking can inform everyday health management in at-risk populations.

Meanwhile, conversational agents and large language models (LLMs) have opened new avenues for personalized health guidance, integrating multimodal inputs such as physical activity, diet, and self-reports to deliver context-sensitive coaching~\cite{10.1145/3706598.3714404,10.1145/3706598.3713819}. Yet despite this progress, CGM data has been largely absent from HCI research on LLM-driven health support. This omission is notable: unlike step counts or diet logs, CGM uniquely captures minute-level physiological responses to everyday behaviors, directly linking lifestyle practices to metabolic health trajectories~\cite{shah2019continuous,kajisa2024correlation}. Without treating CGM as a first-class signal, conversational agents risk overlooking early warning patterns that are invisible in conventional self-tracking. To date, no HCI study has systematically examined how CGM data, when coupled with LLM-based conversational feedback, might scaffold reflection and preventive action in sub-health populations~\cite{jarvis2023continuous}. This gap motivates our autoethnographic investigation.

In May 2025, I experienced a panic attack\cite{Nutt1992Panic} that led to an emergency admission. Although a systematic medical examination revealed no organic pathology, the attending physician emphasized that, as a young researcher, my high work stress, irregular lifestyle, overweight status, and heightened anxiety placed me in a sub-health state requiring immediate adjustment. In response, I undertook a six-week autoethnography. During this period, following medical advice, I tracked my CGM data together with a set of physiological and psychological indicators that can be conveniently captured using common digital health devices. At the same time, I engaged with a chatbot that not only provided personalized reflections and lifestyle recommendations, but also generated plausible explanations for abnormal glucose fluctuations in relation to everyday behaviors. Through this design, I sought to critically examine how CGM-led, data-first multimodal tracking, combined with conversational support, could shape everyday health practices in sub-health contexts. Guided by this motivation, the study addresses the following research question:

\textbf{RQ: How can CGM-led multimodal tracking, supported by large language model--driven conversational agents, shape everyday health management in sub-health contexts?}

By situating this inquiry in the lived experiences of a sub-health researcher, this work contributes to HCI in two main ways. First, it extends CGM research beyond clinical diabetes, providing a rich account of how glucose variability interacts with stress, routines, and lifestyle adjustment in everyday life. Second, it shows how LLM-driven conversational agents can help interpret multimodal health data, supporting reflection and preventive action in populations that are at risk but not yet clinically diagnosed. Together, these contributions advance HCI discussions on personal informatics and the role of AI in preventive self-care.

\section{Related Work}

\subsection{Repositioning CGM: From Disease Management to Everyday Well-Being}

Continuous glucose monitoring (CGM) was first developed for type 1 and type 2 diabetes, offering continuous readings and alerts that transformed self-management~\cite{rodbard2016continuous,danne2017international}. HCI research has examined CGM in this context, including predictive visualizations~\cite{10.1145/3613904.3642234}, integration with wearable data~\cite{10.1145/3706599.3719780}, and smartwatch-based self-management~\cite{10.1145/3706598.3713247}. Recent work has also explored non-invasive prototypes, such as optical and wearable designs, which may make monitoring more ubiquitous, though none have yet achieved medical device–level approval~\cite{leung2025clinical,chang2022highly}. Beyond diagnosed populations, biomedical studies show that non-diabetic individuals exhibit substantial glycemic variability shaped by sleep, stress, and meal timing~\cite{jarvis2023continuous,shah2019continuous}. The \textbf{U.S. FDA} has recently authorized over-the-counter CGM for adults over 18, enabling diet- and exercise-related use~\cite{FDA2024CGM}. Yet HCI has not systematically examined CGM as a preventive tool for at-risk groups such as the overweight or prediabetic.

Responses to food and lifestyle are also highly heterogeneous. Precision nutrition studies reveal wide \emph{inter-individual variability} in postprandial glucose responses, even to identical meals~\cite{hengist2025imprecision,zeevi2015personalized}. Cultural and genetic factors amplify this variability: East Asian high-carbohydrate diets and genetic predispositions contribute to elevated diabetes risk at lower BMI levels compared to Western populations~\cite{cheng2017relevance,nanditha2016diabetes}. This underscores the necessity of personalized interpretation linking lifestyle behaviors with glucose fluctuations, where LLMs can now serve as adaptive “personal nutritionists” that contextualize data into actionable reflections. Commercial platforms such as \textbf{Veri} and \textbf{ChatCGM}~\cite{Veri2024,ChatCGM2025} integrate CGM with AI-based food logging and conversational feedback, but remain oriented toward diabetes management and emphasize food inputs. Little work has examined CGM for preventive health in non-diagnosed populations or its integration with multimodal signals such as sleep, heart rate, or stress. This gap opens opportunities for HCI to re-imagine CGM not only as a clinical management tool but also as a signal for everyday reflection and well-being.

\subsection{Chatbots for Preventive Health and Mental Well-Being}
Conversational agents have been used in HCI to support physical activity change~\cite{jarvis2023continuous,10.1145/3706598.3714404}, mental health~\cite{10.1145/3565698.3565789,10.1145/3554364.3559119}, and habit formation~\cite{10.1145/3613904.3642479} by transforming user inputs, wearables, and dialogue into actionable recommendations. Studies in preventive well-being show that adaptive chatbots delivering positive psychology interventions increase user happiness and resilience through real-time, multi-round interaction. CBT-informed bots also proved effective in reducing stress, insomnia, or depressive symptoms in clinical and non-clinical populations~\cite{Greer2019Use}. Design works further explore how to embed behavior science frameworks, motivate habitual change and support emotional coping over time~\cite{10.1145/3154862.3154914}. While these works highlight strengths such as empathy, adaptivity, and personalized feedback, few of them integrate continuous physiological signals like CGM alongside multimodal data (sleep, heart rate/HRV, diet) for reflection and preventive action in sub-health populations~\cite{jarvis2023continuous}.

\subsection{Autoethnography in HCI and Well-Being}
Autoethnography has become an increasingly prominent method in HCI and well-being research, offering a bridge between personal experience and academic inquiry~\cite{rapp2018autoethnography}. Autoethnography enables deep insights into how individuals interact with technology, particularly in domains such as health, embodiment, and well-being, by allowing researchers to reflect on their own lived experiences and connect them to broader sociocultural contexts~\cite{Kaltenhauser2024Playing}. The approach has been especially valuable in understanding subjective experiences of health and mental well-being, such as managing long COVID, mental distress, and medication, or navigating neurodiversity, by providing nuanced, first-person perspectives that traditional methods may overlook~\cite{Fox2024Autoethnographic,Homewood2023Self-Tracking,Alrøe2025De-centering,10.1145/3706598.3713619,10.1145/3613905.3651096}. While autoethnography is celebrated for its ability to generate rich, authentic data and foster reflexivity, critiques remain regarding its generalizability and scientific rigor, emphasizing the need for clear methodological standards and sociocultural interpretation beyond evocative self-narratives~\cite{Chang2016Autoethnography}. Nevertheless, autoethnography is recognized as a transformative and therapeutic tool, capable of challenging traditional models of expertise and care, and promoting empowerment and co-production in health and HCI contexts~\cite{Custer2022A,Liggins2013Using}.

\section{Methods}

\subsection{Study Design and Procedure}
We conducted a six-week autoethnography in which the first author served as both participant and investigator while living and working in South China.  As the participant, the author is a 24-year-old researcher with a sedentary academic lifestyle, high work stress, and a BMI of 33.8 at baseline, representing a typical case within the broader sub-health group characterized by overweight, irregular routines, and anxiety.
 The study aimed not at testing clinical efficacy but at generating a situated account of how \textbf{Continuous Glucose Monitoring (CGM)}-led multimodal tracking and a chatbot co-shaped everyday health management. To mitigate bias, event thresholds were defined a priori, an audit trail of sensor data and reflections was maintained, and peer debriefing was used. Over 42 consecutive days, daily routines involved continuous passive sensing, event marking (meals, exercise, caffeine, alcohol, naps, and atypical work), and chatbot reflections after salient events. Triggers included postprandial excursions, potential nocturnal lows, days with high glucose variability, low-recovery days (Heart Rate Variability, using Apple Health’s SDNN estimate, thresholded at the participant’s rolling 20th percentile with the baseline recomputed weekly)\cite{wang2012sdnn}, weekend blood-pressure outliers, and user-initiated queries. Weekly measures included body weight, \textbf{Self-Rating Anxiety Scale (SAS)} scores, and resting blood pressure, while major disruptions (e.g., travel, deadlines) were logged for context.

\subsection{Data Sources and Hardware}
We used an arm-worn SIBIONICS CGM (5-min sampling, three consecutive 14-day sensors) alongside an Apple Watch Series 10 paired with an iPhone 15 to record cardiac health (heart rate and HRV), sleep, and daily activity. Periodic measures included weekend blood pressure (Omron HEM-7211), body weight (digital scale), and weekly anxiety assessments using the SAS (Self-Rating Anxiety Scale). All streams except diet were consolidated via Apple Health and uploaded through React Native APIs to the server. Diet was captured as meal photos in Apple Photos, uploaded to the database, and analyzed by a vision-language model (VLM) for food type, portion, and approximate calories/GI. For the first two weeks, before the app front end was complete, chatbot interactions were conducted directly on the server. A physician confirmed these multimodal measures as clinically relevant indicators closely tied to lifestyle.  

\subsection{System and Intervention Logic}
The system comprised both a user-end mobile app and a server-side backend pipeline. On the client side, a dedicated mobile app (A) connected with Apple Health and a custom chatbot interface (B). Multimodal data streams included continuous glucose monitoring (CGM) traces, activities, sleep, and cardiovascular health metrics from the Apple Watch (a–d), as well as blood pressure, body weight, and SAS anxiety scores manually entered into the Health app (e, g). Dietary intake was captured via meal photos and logged through conversations with the chatbot (f). Event-driven data (e.g., meal photos and dialogues) were ingested when created; sensor streams were ingested every five minutes. During analysis, all events were aligned (“snap-to-grid”) to the nearest 5-minute CGM timestamp.

On the backend, incoming health streams were processed through a retrieval-augmented generation (RAG) pipeline. The pipeline maintained an indexed knowledge base on metabolic health, including general mechanisms (e.g., postprandial physiology, sleep and stress effects) and culturally specific food items. When abnormal glucose patterns were detected outside of sleep periods, the backend triggered a chatbot dialogue. Dietary photos were processed by \textbf{Qwen-VL}, a vision–language model that estimated portion size, calories, and approximate glycemic index (GI). These outputs were fused with concurrent physiological indicators and passed to the intelligent agent. The agent used the \textbf{Tencent Hunyuan} large language model to generate context-sensitive feedback, retrieving relevant knowledge snippets when needed and composing explanations, plausible mechanisms, and small actionable suggestions for the user.

The chatbot operated in two modes: scheduled reflective prompts (e.g., 90–120 minutes after meals or ~30 minutes post-exercise) that asked the participant to connect lifestyle context with physiological responses, and on-demand micro-explanations when unexpected sensor fluctuations occurred. Prompt templates emphasized plausible mechanisms, uncertainty, and small actionable experiments such as fiber-first sequencing, short post-meal walks, or adjustments in caffeine and bedtime routines. Safety disclaimers were displayed at all times, diagnostic claims were avoided, and the agent could be muted at the user’s discretion.

\begin{figure*}[t]
    \centering
    \includegraphics[width=1\linewidth]{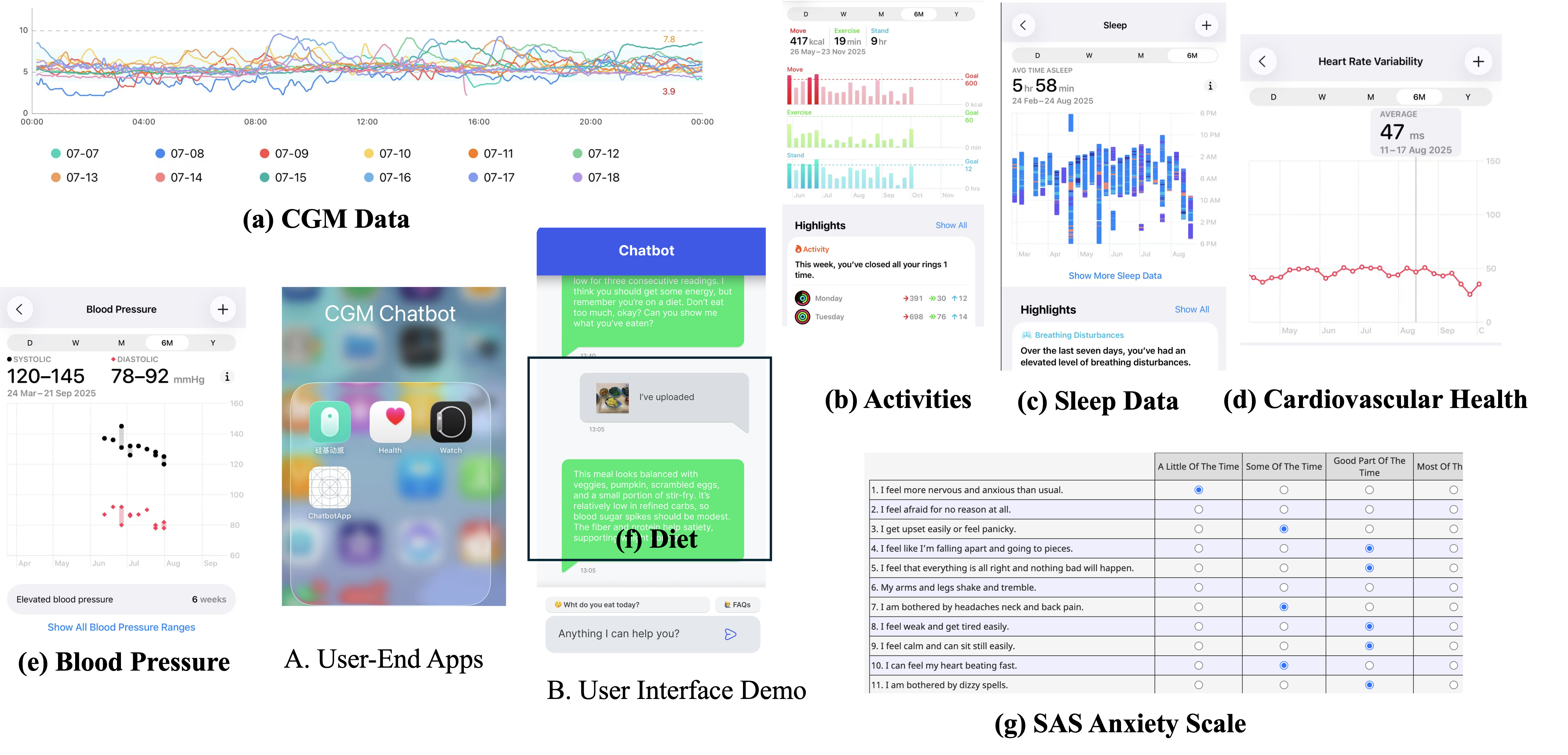}
    \caption{User-end data collection and demo of user interface. The first author tracked personal health using CGM and Apple Watch (a–d). Additional measures including blood pressure, SAS anxiety scores, and body weight were manually entered into the Health app (A) (e, g). Dietary intake was logged via photos and conversations with the chatbot (f). All data, including chatbot interactions, were uploaded at the same frequency as CGM to a private MongoDB database secured with AES-256 encryption.}
    \label{frontend}
\end{figure*}

\begin{figure*}[t]
    \centering
    \includegraphics[width=0.8\linewidth]{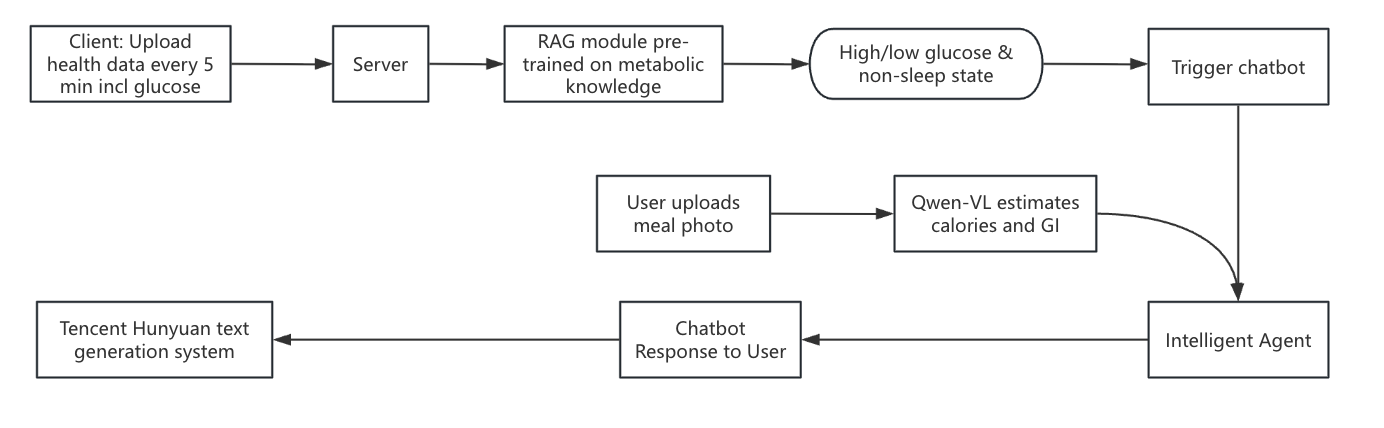}
    \caption{Backend application workflow of the CGM–chatbot system. 
    The client uploads multimodal health data, including continuous glucose monitoring (CGM) values, every five minutes to the server. 
    The server hosts a retrieval-augmented generation (RAG) module pre-trained on metabolic disease knowledge. 
    When abnormal glucose levels are detected outside of sleep periods, a chatbot dialogue is triggered. 
    Users can also upload meal photos, which are processed by Qwen-VL to estimate calories and glycemic index (GI). 
    Both physiological signals and dietary information are passed to the intelligent agent, which integrates them with the Tencent Hunyuan text generation system to produce context-sensitive feedback. 
    The response is then returned to the user as real-time, lifestyle-linked health guidance.}
    \label{fig:backend_workflow}

\end{figure*}

\subsection{Qualitative Corpus and Co-analysis}
The qualitative corpus comprised chatbot transcripts, daily reflections, and contextual annotations (e.g., notes on work routines, exercise, or social meals). The first author conducted open and axial coding to identify recurring functions of the chatbot (e.g., explanation, suggestion, reassurance, review) and corresponding behavioral outcomes (adopted, deferred, or rejected). To enhance credibility, the second author acted as a critical friend: using a week-stratified random scheme, they independently coded approximately 20\% of the corpus, compared interpretations with the first author, and discussed convergence and divergence. This process helped surface blind spots and monitor analytic drift. Disagreements were not forced into consensus but documented in the audit trail with rationale for the adopted categories. We further defined ``index episodes'' for each trigger type as the first qualifying instances, ensuring inclusion of routine as well as atypical examples. Each episode bundled aligned sensor plots, contextual annotations, chatbot responses, and the author’s reflection. A collaborating physician reviewed a subset of these episodes, commenting on plausibility and offering guardrails (e.g., measurement technique, escalation thresholds). This triangulation—across sensors, chatbot, personal reflection, peer coding, and clinical commentary—provided a richer interpretive frame.

\subsection{Analysis and Ethics}
We used an event-aligned, visual analysis: daily metrics were summarized descriptively, and raw CGM trajectories were aligned around meals to compare days differing in sleep, stress, and diet. We looked for repeated motifs (e.g., postprandial spikes, nocturnal lows, morning-after effects) and linked them to chatbot interactions and reflections to pinpoint when feedback timing, wording, or culturally localized substitutions shaped behavior. Design implications followed from these recurrent moments. Data were de-identified and stored in a private AES-256–encrypted MongoDB; the study followed institutional guidance for self-research/low-risk data. The chatbot avoided diagnostic language and provided escalation prompts for persistent anomalies.

Based on institutional guidelines, this autoethnographic self-study was exempt from formal IRB review, as it was considered low-risk and involved only the author as participant.

\section{Findings}

We present four findings from the six-week autoethnography. Each illustrates how CGM-led multimodal tracking, supported by an LLM-driven chatbot, shaped everyday health management. Over this period, the first author and the chatbot engaged in 782 conversational turns, including 21 instances proactively triggered by the chatbot in response to abnormal physiological signals. The findings are organized as event-triggered adjustments, the surfacing of personal sensitivity, multimodal explanations with emotional support, and longitudinal improvements over six weeks.

\subsection{A Midnight Low and a Morning Reminder}

In the third week, my CGM trace recorded a nadir of 2.9 mmol/L during sleep---a low-range value I only noticed the following morning when reviewing the app. Because CGM accuracy can be limited in the hypoglycemic range and \emph{compression lows} are possible during sleep, I could not verify whether this was a true nocturnal hypoglycemia event. Still, the morning reflection prompt from the chatbot highlighted the dip: \emph{``Your glucose dropped into a low range around 2 a.m. This may indicate nocturnal hypoglycemia. Consider adding a small snack before bedtime.''} It further suggested that prolonged fasting windows might not be optimal given my current lifestyle and physiology.  

This reframing transformed what might have remained an invisible or easily dismissed anomaly into a prompt for self-experimentation. Reflecting on my routine, I realized that dinner had been unusually light and I had exercised more than usual. In response, I began adding a small snack (e.g., a whole-grain cracker or a glass of milk) before bedtime on subsequent evenings. Over the next week, my nighttime glucose curves were smoother, with no further dips below the monitoring threshold. This moment underscored how CGM can function not merely as a passive archive of numbers but as a \textbf{sentinel} that surfaces hidden physiological events. Through contextual explanation, the chatbot translated raw data into actionable micro-adjustments, nudging me to change routines in ways that had a tangible effect.

\subsection{The Purple Sweet Potato Betrayal}
I had always believed that purple sweet potato was a low-GI, healthy staple. At one lunch, I ate about \textbf{150 grams} of it. Within thirty minutes, my glucose rose dramatically from \textbf{5.5 to 9.8 mmol/L}—a spike that left me stunned. Confused, I asked the chatbot: \emph{``Why would sweet potato cause such a spike?''} It replied: \emph{``Sweet potatoes can provoke strong responses in some individuals. Your body may be more sensitive to its starches. You might try reducing the portion, or pairing with protein or fat to buffer the effect.''} At first, I dismissed this as a generic line. Yet I experimented over the following days: I cut the portion in half and paired it with eggs or nuts. This time, the postprandial peak stayed below \textbf{7.8 mmol/L}.  

What I took from this episode was not simply that purple sweet potato was ``bad,'' but that my responses could deviate sharply from generalized GI tables. Interestingly, many foods often criticized in East Asian diets as “high GI”—such as white rice or Chinese-style noodles—did not trigger dramatic excursions for me. My postprandial responses to these staples were relatively stable, suggesting that population-level nutritional warnings may not map directly onto my own physiology. The chatbot’s real contribution here was to \textbf{externalize a personal hypothesis}: ``maybe you are sensitive.'' It provided just enough of a scaffold to encourage small experiments, and the results confirmed that my body reacted to purple sweet potato far more strongly than to rice or noodles.  

This realization shifted my relationship to nutrition from passive reliance on universal charts to an iterative cycle of \textbf{personal sensitivity mapping}. Rather than treating all high-GI foods as risky or all low-GI foods as safe, I began to see CGM as a tool for personalized calibration. In this way, the integration of real-time glucose data with conversational feedback highlighted the value of individualized interpretation, revealing patterns invisible to generic dietary advice and enabling more grounded decisions about what to eat in everyday life.

\subsection{Alcohol, Deadlines, and the Comfort of Explanation}
Week four brought a social event where I drank more than usual. The next morning, my blood pressure monitor showed a reading of \textbf{148/95 mmHg}, compared to my typical 130s. Alarmed, I asked the chatbot if something was wrong. Its reply was reassuring: \emph{``Temporary elevation in blood pressure within 24 hours of drinking is common. It can also be linked to poor sleep. Please hydrate and try not to worry.''} The phrasing grounded my experience in a normal physiological response, reducing my anxiety.  

A few days later, during a particularly stressful conference deadline, I experienced an even more severe episode. Having slept only a few hours, my glucose hovered persistently above \textbf{7.8 mmol/L}, my blood pressure exceeded \textbf{150 mmHg}, and my resting heart rate climbed over \textbf{100 bpm}. The sensations were so overwhelming that I briefly considered going to the emergency room. At midnight, I messaged the chatbot: \emph{``I just feel unwell today.''} It responded: \emph{``Your elevated heart rate, glucose, and blood pressure may all be stress responses. This does not mean disease. Try a short breathing or meditation exercise, and sip some electrolyte water to stay balanced.''} I followed its guidance, repeating several rounds of paced breathing and meditation while hydrating. Gradually, my heart rate settled and the sense of panic subsided, even though glucose remained moderately high.  

What stood out was that the chatbot did not over-medicalize the data; instead, it offered \textbf{credible, multifactorial attributions} and paired them with small, actionable calming techniques. This combination of explanation, reassurance, and practical coping suggestions reframed ambiguous and alarming sensations into something understandable, offering \textbf{anxiety-aware coaching} at moments when I needed it most.

\subsection{Six Weeks of Gradual Change}

\begin{figure*}[t]
    \centering
    \includegraphics[width=1\linewidth]{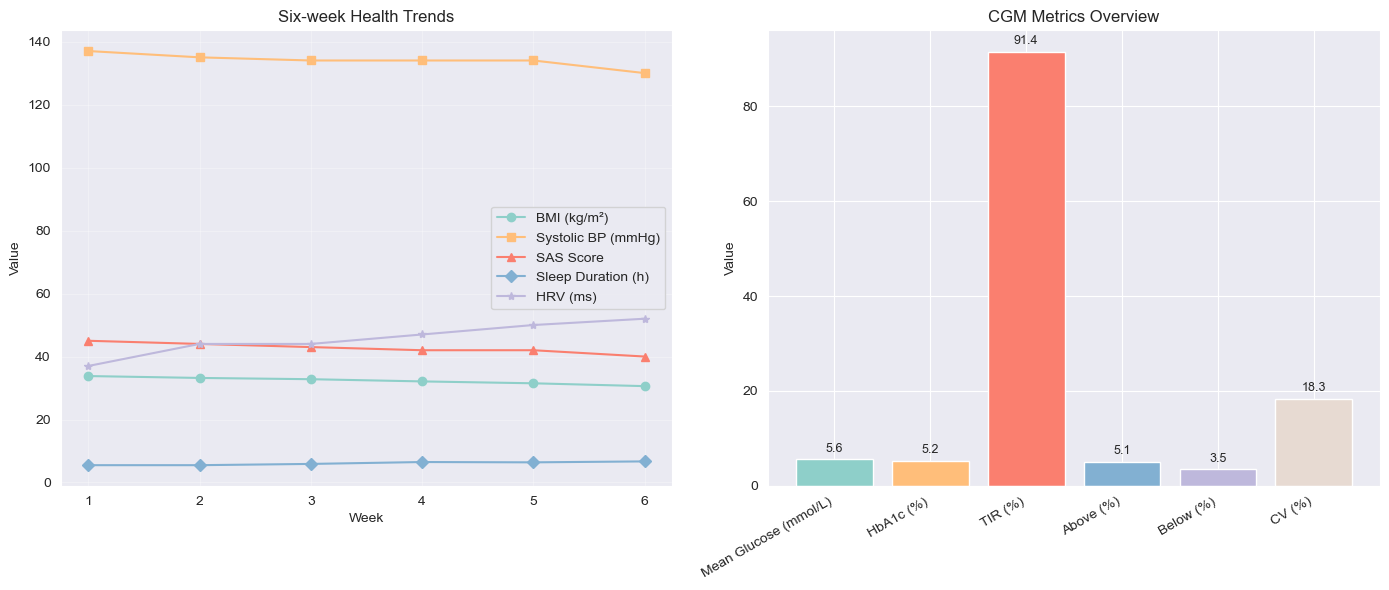}
    \caption{Six-week health metrics overview. Left: Longitudinal trends of BMI (kg/m²), systolic blood pressure (mmHg), Self-Rating Anxiety Scale (SAS) scores, average sleep duration (hours), and heart rate variability (HRV, ms). Across the six weeks, BMI, blood pressure, and SAS scores decreased, while sleep duration and HRV increased, indicating improved physiological and psychological regulation. Right: Summary of CGM-derived metrics, including mean glucose, estimated HbA1c, time in range (TIR), excursions above and below range, and coefficient of variation (CV). These metrics remained largely within healthy ranges, highlighting the visibility of lifestyle impacts without evidence of insulin resistance.}
    \label{fig:health}
\end{figure*}

While the first three findings captured episodic adjustments, the most profound realization emerged at the end of six weeks, when cumulative trends came into view. My \textbf{BMI decreased steadily from 33.8 to 30.6}, marking a clinically meaningful reduction. Average \textbf{systolic blood pressure dropped from 137 mmHg to 130 mmHg}, with weekend fluctuations becoming less erratic. My \textbf{Self-Rating Anxiety Scale (SAS) scores fell from 45 to 40}, corresponding to an improved subjective sense of coping. Meanwhile, restorative dimensions showed upward progress: \textbf{average nightly sleep lengthened from 5.5 to 6.7 hours}, and \textbf{heart rate variability (HRV) increased from 37 ms to 52 ms}, both indicating improved recovery capacity.

From a glycemic perspective, my \textbf{mean glucose level was 5.6 mmol/L}, with an estimated \textbf{HbA1c of 5.2\%}. \textbf{Time in Range (3.9–7.8 mmol/L) was 91.4\%}, with \textbf{5.1\% above range and 3.5\% below range}; the \textbf{coefficient of variation was 18.3\%}.\footnote{These glycemic metrics were automatically computed by the CGM system. While such indicators are typically used in diabetes management for assessing disease progression, here they are presented only to demonstrate that overall glucose stability remained relatively good.}According to a consulting physician, these readings do not approach thresholds of insulin resistance. More importantly, the observed reductions in \textbf{BMI, blood pressure, and anxiety}, alongside increases in \textbf{sleep and HRV}, represent valuable improvements for long-term health. At the same time, CGM made visible how even small shifts in meals, sleep, or stress translated into immediate glucose dynamics, underscoring the tangible impact of daily lifestyle choices.

This continuous feedback, coupled with the chatbot’s contextual explanations and nudges, helped me reframe what it means to manage health as an overweight young researcher living in a sub-health state. Rather than chasing a “magic solution,” I began to see self-care as an ongoing dialogue—between data and self, between everyday routines and broader aspirations. The changes I observed should not be taken as causal evidence, but as short-term signals moving in the same direction as ongoing lifestyle adjustments. These aligned trends suggested that embedding CGM into conversational support may scaffold healthier routines and strengthen resilience, even within a limited observational window. Over time, this process highlighted how one can sustain incremental improvement and pursue healthier living and healthier research simultaneously. Beyond individual benefit, this experience underscored how technology-mediated reflection can illuminate the subtle links between body, lifestyle, and work in ways that static guidelines often cannot.

\section{Discussions and Future Work}
\subsection{Answering the Research Question}
Taken together, our four findings demonstrate that CGM-led multimodal tracking, when coupled with an LLM-driven chatbot, reshaped everyday health management in sub-health contexts through three interlocking mechanisms. First, CGM surfaced hidden physiological dynamics—such as nocturnal lows—that would otherwise remain invisible. Second, conversational feedback translated raw data into hypotheses and actionable micro-adjustments, allowing the user to iteratively map personal sensitivities beyond population-level nutritional or lifestyle guidelines. Third, multimodal explanations integrated physiology, behavior, and emotion, offering anxiety-aware coaching that reframed alarming sensations into manageable experiences. Over six weeks, these episodic insights accumulated into measurable improvements in weight, blood pressure, anxiety, and sleep. In this way, health management shifted from static adherence to universal recommendations toward a dynamic dialogue between data and self, suggesting that technology-mediated reflection can scaffold resilience and sustainable lifestyle change in sub-health conditions.

\subsection{From CGM to CXM: Continuous Experience Monitoring}
While CGM offered a powerful lens on glycemic dynamics, the study also points to the value of extending this paradigm beyond glucose. We propose the notion of \textbf{Continuous Experience Monitoring (\textbf{CXM})}, where multimodal sensor streams—heart rate, HRV, sleep, blood pressure, activity, and even digital traces—are integrated into a longitudinal narrative of everyday health~\cite{nia2015energy}. Unlike episodic checkups or static guidelines, \textbf{CXM} emphasizes ongoing, situated feedback loops between physiological signals, lived routines, and reflective interpretation. Such a framework acknowledges that sub-health is rarely captured by a single metric, but emerges from complex interactions of diet, stress, sleep, and social life. 

Importantly, this direction resonates with an ongoing industrial trend: many technology companies are already embedding additional physiological measures such as blood pressure monitoring~\cite{kario2020first}, sleep apnea detection~\cite{hayano2020quantitative}, and continuous respiratory tracking into their consumer-grade wearables. These developments suggest that \textbf{CXM} is not only conceptually compelling but also technologically and commercially viable. By anchoring health management in continuous, contextualized experience rather than isolated readings, \textbf{CXM} thus represents an inevitable trajectory for future health monitoring—opening pathways for more personalized, adaptive, and resilient forms of self-care.

\subsection{Chatbots as Health Companions: Benefits and Ethical Concerns}

The role of the LLM-driven chatbot in this study highlights both promise and tension. On the one hand, chatbots provided immediate, empathetic, and context-sensitive feedback at moments when professional medical guidance was unavailable, offering reassurance, explanation, and micro-interventions that helped reduce anxiety and sustain behavior change~\cite{Wah2025Revolutionizing,Manole2024An}. This suggests that conversational agents can serve as accessible health companions, particularly for populations navigating sub-health conditions without requiring constant clinical attention~\cite{Laymouna2024Roles}. On the other hand, their deployment raises critical ethical questions.~\cite{Parviainen2021Chatbot} Chatbots may inadvertently blur the line between coaching and diagnosis, creating risks if users over-rely on non-clinical advice~\cite{Mitchell2021Automated}. Issues of data privacy, algorithmic transparency, and cultural sensitivity further complicate their role in sensitive health contexts~\cite{Fan2020Utilization}. Designing chatbot systems for health thus requires not only technical accuracy but also careful governance: mechanisms for safeguarding data, clear boundaries between support and clinical care, and participatory approaches that respect users’ autonomy and well-being~\cite{10.1145/3706598.3713593}.

\subsection{Cultural and Contextual Sensitivity}

Our experience highlights that health technologies are never neutral instruments but are deeply embedded in cultural practices, values, and social contexts. In the Chinese setting, dietary norms (e.g., high reliance on refined carbohydrates)~\cite{yu2013dietary}, traditional “wellness culture,”~\cite{nestler2002traditional} long working hours~\cite{chu2021impact}, and family expectations shape how individuals perceive and act upon “healthy advice.”~\cite{campbell1998diet} In our case, the episode of the \textbf{“purple sweet potato betrayal”} illustrated how cultural framings of foods as “healthy staples” may diverge from individual physiological responses. Although purple sweet potato is widely regarded as a nutritious low-GI food, CGM revealed a strong postprandial excursion for the author. This underscores how cultural categories of health and nutrition coexist with personalized variability, and how technology-mediated reflection can bridge the two.

Beyond diet, broader socio-structural realities in China—such as uneven distribution of medical resources, rural–urban disparities, and heterogeneous health beliefs—further shape the feasibility and trustworthiness of digital health interventions~\cite{Huang2017Beyond}. Prior work has shown that culturally diverse populations interpret and respond to digital health recommendations differently~\cite{Whitehead2022Barriers}. If conversational agents ignore such contextual differences—such as language registers, colloquial expressions, or local beliefs—they risk producing misaligned or even alienating advice. 

We therefore argue that future CGM–chatbot systems should incorporate \textit{cultural adaptation modules}. These may adjust feedback tone, explanatory framing, and intervention strategies according to users’ cultural backgrounds, dietary patterns, or linguistic preferences. Such sensitivity would enhance resonance and uptake across diverse contexts, making CGM-driven conversational systems not only technically accurate but also culturally attuned health companions. Designers should therefore incorporate participatory and co-design processes with local users to ensure that language tone, dietary framing, and behavioral nudges align with culturally specific values and routines.

\subsection{Limitations and Future Work}
As an autoethnographic study, our findings are inherently limited by the sample size of one and the situated positionality of the first author. Rather than providing generalizable evidence, the study should be read as a prospective probe that foregrounds how CGM-led multimodal tracking and chatbot feedback can be woven into everyday sub-health management. Second, the current system is highly dependent on Apple’s mobile ecosystem (iOS and HealthKit), which constrains portability and limits accessibility for users across diverse platforms. Finally, while our prototype demonstrated feasibility, future work will extend the system in three directions: (1) enhancing functionality, personalization, and data privacy safeguards to mitigate ethical risks; (2) broadening platform compatibility for wider adoption; and (3) conducting larger-scale user studies to examine whether CGM-led multimodal monitoring and chatbot-based reflection can help broader sub-health populations regain healthier routines. It is also important to note that current chatbots are designed only for preventive health support rather than clinical diagnosis; when users experience persistent or severe physical discomfort, the system should proactively remind them to seek professional medical care.

\section{Conclusion}
This autoethnographic study explored how Continuous Glucose Monitoring (CGM), when coupled with chatbot-based reflection and multimodal sensing, can reshape everyday health management in sub-health conditions. Over six weeks, the first author’s lived experience showed three interlocking mechanisms: CGM surfaced hidden physiological dynamics; chatbot conversations translated raw signals into hypotheses and micro-adjustments; and the integration of multimodal data reframed alarming episodes into constructive insights. Together, these mechanisms illustrate how data-led, dialogue-driven systems can scaffold resilience and foster sustainable lifestyle changes. While the single-participant design limits generalizability and the system currently depends on Apple’s ecosystem, the study offers a situated probe into how CGM-led multimodal tracking and conversational coaching may extend preventive health beyond static guidelines. Building on this, we propose the broader paradigm of Continuous Experience Monitoring (\textbf{CXM}), in which physiological, behavioral, and digital traces are continuously interpreted through interactive feedback loops. Future work will enhance personalization, privacy safeguards, and platform portability, and involve larger-scale evaluations to assess whether CGM plus chatbot support can help wider sub-health populations return to healthier trajectories. By highlighting the value of coupling real-time data with conversational interpretation, this work points toward more adaptive, contextualized, and humane approaches to technology-supported self-care.

\section*{Disclosure about Use of LLMs}
We used the built-in AI assistant in \textbf{WebStorm} to support prototype development in \textbf{Node.js} and \textbf{React Native (TypeScript)}, as well as to assist in writing a \textbf{Swift-to-React Native bridge} for accessing \textbf{HealthKit} data. In addition, we employed \textbf{GPT-5} for limited language refinement in non-substantive parts of this paper. \textit{LLMs did not contribute to the conceptualization, study design, or generation of creative content.}

\section*{Acknowledgments}

We would like to express my deepest gratitude to the Red Bird MPhil Program at the Hong Kong University of Science and Technology (Guangzhou) for providing me with generous support, resources, and funding, which have been instrumental in the successful completion of our research.

We sincerely thank the medical professionals from \textbf{Yancheng No.1 People's Hospital} and \textbf{Beijing Anding Hospital, Capital Medical University} for their valuable medical guidance and educational support throughout this study. Their expertise in endocrinology and mental health provided essential insights that informed the self-monitoring protocol and the interpretation of physiological and psychological data.

\bibliographystyle{ACM-Reference-Format}
\bibliography{refs}

\end{document}